\documentclass[aps,pre,reprint,groupedaddress]{revtex4-1}

\usepackage{graphicx}
\usepackage{float}
\usepackage{epstopdf}

\begin{document}

\title{Nonlinear stability of laboratory quasi-Keplerian flows}

\author{E. M. Edlund and H. Ji}
\affiliation{Princeton Plasma Physics Laboratory, Princeton, New Jersey 08543, USA}
\thanks{Accepted for publication in PRE.}

\date{\today}

\begin{abstract}
Experiments in a modified Taylor-Couette device, spanning Reynolds numbers of $10^5$ to greater than $10^6$, reveal the nonlinear stability of astrophysically-relevant flows. Nearly ideal rotation, expected in the absence of axial boundaries, is achieved for a narrow range of operating parameters. Departures from optimal control parameters identify centrifugal instability of boundary layers as the primary source of turbulence observed in former experiments. By driving turbulence from a series of jets we demonstrate the robustly quiescent nature of quasi-Keplerian flows, indicating that sustained turbulence does not exist.
\end{abstract}

\pacs{47.32.Ef, 42.27.Jv, 42.27.N-, 47.20.lb}

\maketitle

The sheared flows of dust, gas and plasma in accretion disks are reservoirs of free energy with the potential to drive turbulence that enhances the outward transport of angular momentum. While theoretical studies indicate that magnetic fields play an important role in hot, well-ionized accretion disks through the magneto-rotational instability (MRI) \cite{BH}, colder accretion disks, like weakly-ionized proto-planetary systems, are stable with respect to infinitesimal perturbations even in the limit of vanishing viscosity. Various mechanisms for enhanced transport in the hydrodynamic regions have been proposed, including coupling to magnetohydrodynamic surface layers \cite{Gammie,Bai}, incompressible effects such as baroclinic instability \cite{Klahr} and Rossby wave instability \cite{Meheut}, or through a subcritical transition to turbulence \cite{Richard}. This last mechanism, that is, sustained turbulence triggered in laminar flow by finite-amplitude perturbations, has been observed in rectilinear systems such as plane Couette flow \cite{Daviaud} and pipe-flow \cite{Hof_prl} and remains a candidate to explain enhanced transport in cold accretion disks. Prior experiments and simulations fall on both sides of the debate as to whether there exists a purely hydrodynamic pathway to sustained turbulence in rotating fluid systems like those of accretion disks. We demonstrate through a series of experiments that turbulence from boundary layers can obscure the inherent stability of the bulk flow in laboratory experiments, and that remarkably quiescent, robustly stable flows develop when unstable boundary layers are mitigated, despite applications of large perturbations. These observations suggest that additional physics beyond that of incompressible hydrodynamics is necessary for enhanced angular momentum transport in accretion disks.

\begin{figure}[t]
\includegraphics[width=8cm]{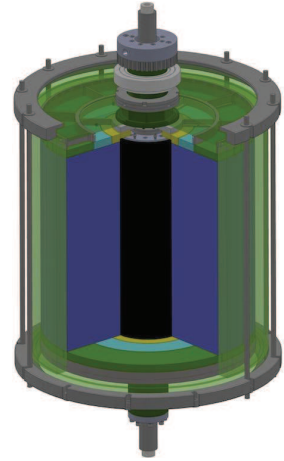}
\caption{(color online) Drawing of the HTX device illustrating the segmented axial boundaries: the inner-most element (yellow) is connected to the inner cylinder (black), the   -most element (green) is connected to the outer cylinder (green) and the middle components (blue) are rings that can rotate differentially with respect to the cylinders.}
\end{figure}

Synopses of recent discussions regarding the laboratory approach to studies of quasi-Keplerian flows, that is, flows with $d \left| \Omega \right| / dr < 0$ and $d \left| r^2 \Omega  \right| / dr > 0$, can be found in {Refs.} \cite{Ji_Balbus_PT} and \cite{Balbus_nature}. The primary experimental apparatus used for such studies is the Taylor-Couette device, in essence two co-axial cylinders capable of differential rotation with the working fluid between (see Fig.~1). A rich space of secondary motions exists when Taylor-Couette devices are operated at low Reynolds numbers of order $10^3$ ($\text{Re} \sim vL/\nu$, is the ratio of inertial to viscous forces, where $v$ is a representative system velocity, $L$ a system length-scale and $\nu$ is the kinematic viscosity), with bifurcations and hysteresis between states \cite{Coles,Andereck}. Subcritical behavior has been observed in linearly stable, rotating flow \cite{Wendt33, Taylor36, Borrero, Burin12}, but in the astrophysically irrelevant regime of $d \left| \Omega \right| / dr > 0$. Using  a Taylor-Couette apparatus with modified axial boundaries, Ji \textit{et al.} \cite{Ji_nature, SchartmanAA} found large $\text{Re}$, quasi-Keplerian flows to have very low levels of fluctuations and measured the local transport of angular momentum to be orders of magnitude too small for astrophysical relevance. The latter paper \cite{SchartmanAA} in particular, claims that a subcritical transition likely does not exist, but does not prove this point directly with finite amplitude perturbations. In contrast, Paoletti and Lathrop ({P\&L}) \cite{Paoletti}, using a Taylor-Couette device of the classical configuration, where the axial boundaries co-rotate with the outer cylinder, found enhanced torque on the inner cylinder in the quasi-Keplerian regime of operation and hypothesized that this was due to a sustained turbulent state, the result of hysteresis, and by implication, a sign of a subcritical transition to turbulence. However, the {P\&L} studies did not directly measure the rotation profiles and therefore could not distinguish the stability properties of the bulk flow from that of the boundary layers.

Numerical simulations of the Ji \textit{et al.} and {P\&L} experiments conducted by Avila \cite{Avila} attempted to reconcile differences in measured angular momentum transport. Avila concluded that the enhanced torque in the quasi-Keplerian regime of the {P\&L} experiments arose from boundary-layer turbulence near the inner cylinder. A highly turbulent state was also found for the Ji \textit{et al.} experiments, contrary to observations of quiescent flows at large $\text{Re}$, though this may be consistent with lower-$\text{Re}$ experiments in the range of $10^4$ to $10^5$ where enhanced Reynolds stress was measured \cite{Ji_nature}. Simulating accretion disk geometries, Balbus, Hawley and Stone (BHS) \cite{BHS} performed three-dimensional, finite volume simulations for $\text{Re}$ in the range of $10^3$ to $10^4$  and found that hydrodynamic instability is possible only when the rotational shear exceeds the limit for centrifugal stability. However, at lower values of rotational shear, for astrophysically-relevant flows, BHS found only dissipation of the seed turbulence. In contrast, shearing box simulations have found both outward transport of angular momentum \cite{Lesur_Ogilvie} and, in the simulations of Lesur and Longaretti ({L\&L}) \cite{LL}, sustained turbulence based on extrapolations of turbulence, with critical $\text{Re}$ for transition spanning the range of $10^6$ to $10^{26}$. In short, there is little agreement among numerical simulations as to whether subcritical turbulence exists, and if it does, what $\text{Re}$ is needed for transition. Our studies seek to resolve this controversy and guide future numerical studies by demonstrating the nonlinear stability of quasi-Keplerian flows.

\begin{table}[t]
\caption{Physical parameters of the HTX device.}
\label{params}
\begin{ruledtabular}
\begin{tabular}{l  c  r}
Parameter & Symbol & Value \\ \hline
Inner cylinder (IC) radius & $r_1$ & $6.9$ cm \\
Outer cylinder (OC) radius & $r_2$ & $20.3$ cm \\
Ring inner radius & $r_3$ & $7.9$ cm \\
Ring outer radius & $r_4$ & $13.4$ cm \\
Height & $h$ & $39.7$ cm \\
Aspect ratio, $h/(r_2 - r_1)$ & $\eta$ & 2.96 \\
IC angular speed & $\Omega_1$ & 0-1400 RPM \\
OC angular speed & $\Omega_2$ & 0-400 RPM \\
Ring angular speed & $\Omega_3$ & 0-700 RPM \\
\end{tabular} 
\end{ruledtabular}
\end{table}

The apparatus used for the experiments reported here is the Princeton Hydrodynamic Turbulence Experiment (HTX), a Taylor-Couette design modified to includeaxial boundaries that are split into multiple components with independently controllable rings (see Fig.~1 and Table~1). The axial boundaries are acrylic to allow for optical diagnostic access, with radial gaps of $0.2$ cm between the various rotating components. The working fluid for these studies is water, with a kinematic viscosity of approximately $1 \times 10^{-6}$ m$^2$/s. A laser-Doppler-velocimetry (LDV) diagnostic is mounted on a radial traverse, viewing upward through the base of the device, and measures the speed of reflective tracer particles. Calibration of the LDV system is performed by measurement of solid body rotation, which sets the noise floor at approximately $0.5$\% of the mean velocity. An array of $16$ radially oriented nozzles mounted on the IC provide perturbations to the flow. Eight nozzles are placed at the mid-plane and four each at a distance of $4.5$ cm from each axial boundary. Fluid is circulated through the experimental volume via a bisected, hollow IC axle, driven by an externally mounted, three-stage centrifugal pump that delivers a maximum flow of $9 \times 10^2$ cm$^3$/s.

\begin{figure}
\includegraphics[width=8cm]{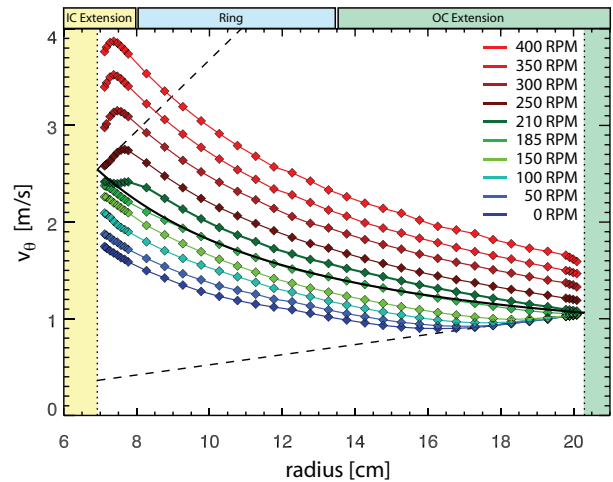}
\caption{(color online) A subset of the measured $v_\theta$ profiles as a function of ring speed for $\Omega_1 = 350$ RPM and $\Omega_2=50$ RPM, with dashed lines indicating the solid body rotation profiles at these speeds. The solid black trace shows the Couette profile. Statistical uncertainties in the mean velocity measurements are less than $0.5\%$.}
\end{figure}

The laminar rotation profiles of Taylor-Couette devices (in the absence of infleunce from axial boundaries) are determined by the condition of constant radial flux of angular momentum, resulting in ideal Couette rotation, profiles of the form $\Omega_C = \Omega_a + \Omega_b \left( r_g / r \right)^2$, where $r_g= \sqrt{r_1 r_2}$. The parameters $\Omega_a =\left( r_2^2 \Omega_2 - r_1^2 \Omega_1 \right) / \left(r_2^2 - r_1^2 \right)$ and $\Omega_b =  r_g^2 \left(\Omega_1 - \Omega_2 \right) / \left(r_2^2 - r_1^2 \right)$ represent the contributions of solid body and irrotational flow, respectively. The global dimensionless shear is defined for these systems by the radial average of $q = - d \ln \Omega / d \ln r$. Rotation in accretion disks nearly follows the Keplerian form, $v_\theta \propto r^{-1/2}$, on account of the small radial pressure gradients and mass typical of these systems, and corresponds to $q \approx 3/2$. Quasi-Keplerian flows are defined by $0<q<2$ and centrifugal (Rayleigh) instability occurs in flows with $q>2$ in the large $\text{Re}$ limit. This work focuses on studies performed at $q=1.8$ where the closest match to Couette rotation is achieved while maintaining local centrifugal stability of the Coutte profile. For the experiments relevant to the discussions of Figs.~2-4, the inner cylinder (IC) speed was held at $350$ RPM and the outer cylinder (OC) at $50$ RPM. The shear Reynolds number, $\text{Re}_s$, of this case is $5 \times 10^5$, where $\text{Re}_s= \Delta \Omega \: \Delta r \: \bar{r} / \nu$, $\Delta \Omega = \Omega_1 - \Omega_2$, $\Delta r = r_2 - r_1$, $\bar{r} = \left(r_1 + r_2 \right) /2$ and $\nu$ is the kinematic velocity. With the IC and OC speeds thus fixed, the ring speed was varied from $0$ RPM to $400$ RPM in $10$ RPM  increments, except in the vicinity of $185$ RPM where $5$ RPM increments were used.

\begin{figure}[t]
\includegraphics[width=8cm]{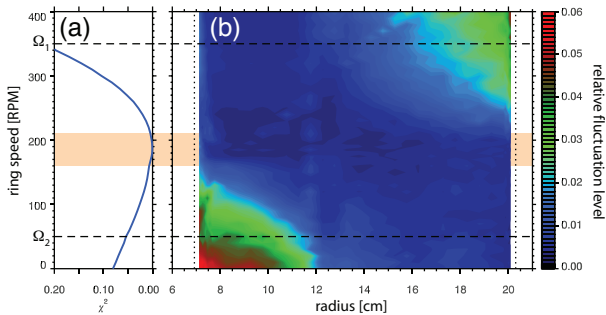}
\caption{(color online) (a) $\chi^2$ values for the experimental $v_\theta$ profiles relative to the Couette profile. (b) The measured fluctuation in $v_\theta$, normalized to the mean $v_\theta$ for each ring speed and radial location. The beige band indicates the range of optimal performance.}
\end{figure}

The profiles of mean $v_\theta$ presented in Fig.~2 demonstrate the sensitivity of the global flow to the influences of the axial boundary conditions. These measurements were conducted at the midplane, and measurements in this device and others \cite{SchartmanRSI} indicate that the profiles exhibit axial uniformity to within a few centimeters of the boundaries. When the rings rotate in lock-step with the OC at $50$ RPM, the axial boundary conditions are very nearly that of the classical Taylor-Couette configuration, excepting the $1$ cm extension from the IC. Under these conditions the measured azimuthal velocities are substantially less than the Couette profile (second curve from the bottom in Fig.~2), with a large jump in velocity occurring over a boundary layer near the IC. At the highest ring speeds, the measured velocities are substantially larger than Couette, and result in boundary layers at both the IC and OC, though only the outer one is centrifugally unstable. When the ring speed is within approximately $\pm 25$ RPM of $185$ RPM we find excellent agreement between the experimental and ideal profiles. Scaling of the overall system Reynolds number indicates that there is little change in the shape of the mean velocity profiles when the system is operated near optimal ring speed. However, when the ring speed is either substantially higher or lower than this optimal range, the profiles of the mean velocity progressively deviate from the Couette solution as the Reynolds number is increased, an effect attributed to increasing agreement with the Taylor-Proudman theorem as the Ekman number ($\text{E}=\nu/ \Omega r^2$) is decreased \cite{HF}. The magnitude and radial extent of fluctuations are strongly correlated with large negative values of the velocity gradient of the boundary layers, as can be seen in the spatial distributions of velocity fluctuations as a function of ring speed in Fig.~3. The transition from the bulk flow to the boundary speeds set by the IC and OC occurs in Stewartson boundary layers (SBLs), whose thickness scales like $\text{E}^{1/4}$ \cite{Stewartson}. Prior experiments have shown that SBLs can become turbulent at sufficiently high $\text{Re}$ \cite{Fruh,Baker67,Hart}.

\begin{figure}[t]
\includegraphics[width=8cm]{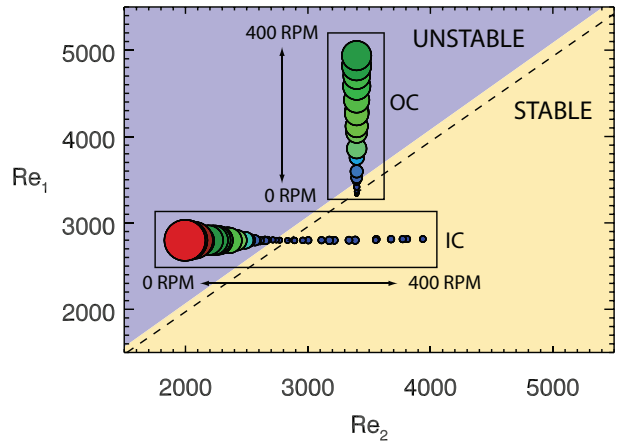}
\caption{(color online) Measurements of near-wall fluctuations are presented with circle radius and color proportional to fluctuation amplitude, using the same color scale of Fig.~3b. Transitions from quiescent states (blue) to increasing fluctuations (green to red) coincide with passing of the Taylor marginal stability boundary. The dashed line is solid body rotation.}
\end{figure}

Inertial effects reduce the thickness of SBLs by roughly a factor of three when the parameter $\mu = \text{E}^{1/2} / \text{Ro}^2$ is of order one or less, where $\text{Ro} = \Delta \Omega/ \Omega$ is the Rossby number of the boundary layer \cite{BH73}. The regions of large gradients in angular velocity near the radial boundaries in the experiment have $\mu \approx 0.1$, suggesting that the SBL thickness should be modeled by the form $\delta r_{SBL} = r \text{E}^{1/4}/3$, corresponding to estimates of boundary layer thicknesses of  $0.1$ cm and $0.3$ cm at the IC and OC, respectively. Defining shear Reynolds numbers for the inner and outer limits of the SBL near the IC, for example, $\text{Re}_1 = v_1 \delta r_{SBL}/\nu$ and $\text{Re}_2 = v_{fluid} \; \delta r_{SBL}/\nu$, where $v_1$ is the linear speed of the IC and $v_{fluid}$ is the measured fluid velocity just outside of the boundary layer, the observed stability of the SBLs is found to be in good agreement with the Taylor marginal stability criterion \cite{Taylor23} for centrifugal instability in low-$\text{Re}$ flows (see Fig.~4). It should be cautioned, however, that Taylor's analysis considered flows bounded by two solid cylinders, whereas the SBLs are bounded on only one side. Nonetheless, the presence of SBLs and the local nature of the centrifugal instability suggest that boundary-driven turbulence should be ubiquitous in the limit of small Ekman numbers if the axial boundaries are not sufficiently optimized as demonstrated here.

\begin{figure}[t]
\includegraphics[width=8cm]{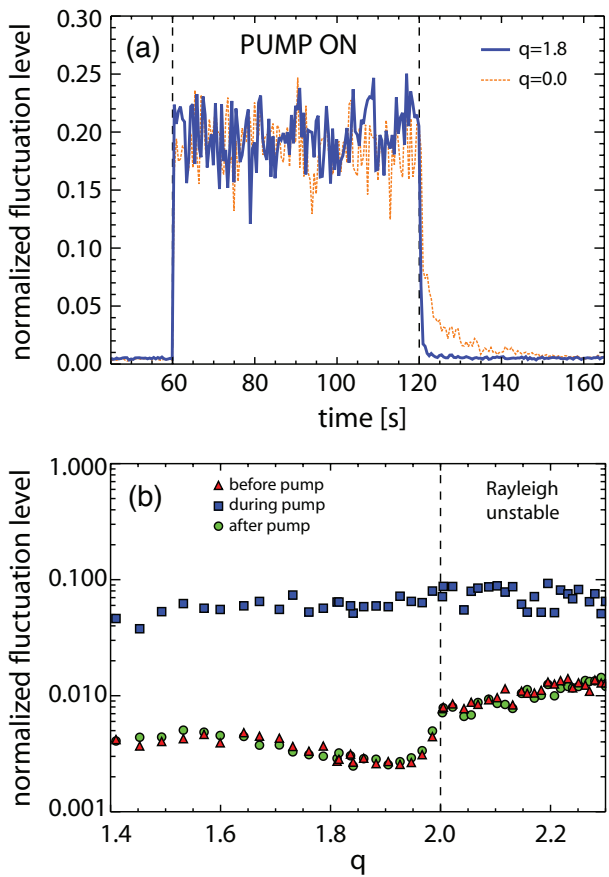}
\caption{(color online) Measurements of relative $v_\theta$ fluctuations for (a) $m=0$ perturbations into $q=1.8$ (blue) and $100$ RPM solid body rotation (orange) flows, and (b) $m=2$ perturbations into a range of flows of different $q$ values.}
\end{figure}

By avoiding regimes of operation where boundary layers produce turbulence that propagates well into the bulk flow, the stability of true quasi-Keplerian flows can be established with respect to finite amplitude perturbations. Multiple methods of perturbation, including fluid jets, pulsing of the IC, OC and the rings, have been applied to these flows, and over the course of thousands of experiments none have exhibited a subcritical transition; quasi-Keplerian flows always return to the basal fluctuation level. Figure 5a shows the temporal response to forced turbulence from the IC nozzle array. These forced fluctuations span about a third of the gap, are localized with a FWHM axial extent of approximately $1$ cm, and are associated with a nearly vertically uniform decrease in the mean velocity of roughly 5\% at the highest pump flow rates. The induced fluctuations decay slowly in solid body flow over roughly $100$ rotation periods, in contrast to the sheared flow where they damp out in just a few rotation periods. These results are qualitatively consistent with recent simulations of periodic cylinders, indicating that turbulence induced from a Rayleigh-unstable initial state decays as the boundary conditions transition to the quasi-Keplerian regime \cite{Monico}, but with much longer lifetimes than are observed in our experiments. Figure 5b compares the relative fluctuation levels before, during and following application of perturbations for a range of $q$ over which there is no substantive difference in fluctuation levels before and after application of the perturbations, indicating a strong stability of quasi-Keplerian flows. The rise in basal fluctuation level slightly below $q=2$ is due to unstable boundary layers near the IC, attributed to a sub-optimal geometry of the axial boundaries for this regime of operation. The enormous reduction in dissipation time-scale for turbulence in quasi-Keplerian flows, relative to dissipation under solid body rotation, suggests that turbulent structures are sheared apart before they can significantly enhance angular momentum transport. No significant difference in turbulence lifetime was observed for jets arranged in vertically asymmetric, azimuthally symmetric ($m=0$) and vertically symmetric, azimuthally asymmetric ($m=2$) configurations.  Experiments have been conducted with similar profiles at $\text{Re}$ up to $2 \times 10^6$, just beyond the linear extrapolation threshold predicted by the {L\&L} simulations \cite{LL}, albeit at a relative fluctuation level of approximately 1\%, with no indication of a subcritical transition under any condition.

The use of segmented axial boundaries is shown to optimize performance in terms of creating rotation profiles that closely match the Couette profile and have very low levels of inherent fluctuations. When the system is operated in the classical configuration we observe high levels of fluctuations that originate from boundary layers near the IC. This turbulence can propagate well beyond the boundary layer and may falsely give the impression of instability of the bulk flow. It is interesting, and not understood, why the turbulence generated near the IC at the low ring speeds of these studies appears to saturate and be confined to the inner half of the gap. Seed turbulent states from externally forced perturbations, that in the context of astrophysical systems are very large, have revealed a robust stability over a broad range in dimensionless shear. Based on these observations, it appears that a subcritical transition to sustained turbulence does not occur for laboratory quasi-Keplerian flows at Reynolds numbers less than $2 \times 10^6$.

\bibliography{Edlund_arxiv}

\end{document}